\documentclass[journal]{IEEEtranTCOM}

\IEEEoverridecommandlockouts 
\usepackage{epsfig,latexsym}
\usepackage{float}
\usepackage{indentfirst}
\usepackage{amsmath}
\usepackage{amssymb}
\usepackage{cite}
\usepackage{bm}
\usepackage{url}
\usepackage{lastpage}
\usepackage{color}
\usepackage{fancyhdr}
\usepackage{multirow}
\usepackage[square, comma, sort&compress, numbers]{natbib} 
\usepackage{setspace}
\normalsize

\renewcommand\appendixname{Appendix}

\begin{document}
\title{Design of Polar Codes with Single and Multi-Carrier Modulation on Impulsive Noise Channels using Density Evolution}

\author{Zhen Mei, Bin Dai, Martin Johnston,~\IEEEmembership{Member,~IEEE} and Rolando Carrasco

\thanks{Zhen Mei is with the Science and Maths Cluster,  Singapore University of Technology and Design, Singapore, 487372 (e-mail: mei\_zhen@sutd.edu.sg).}
\thanks{Bin Dai is with the School of Electronic and
Information Engineering, Beihang University, Beijing, 100191, China (e-mail: daibinok@buaa.edu.cn).}
\thanks{Martin Johnston and Rolando Carrasco are with the School of Engineering, Newcastle University, Newcastle-upon-Tyne, UK (e-mail:  martin.johnston@ncl.ac.uk, r.carrasco@ncl.ac.uk).}
} 


\maketitle
\begin{abstract}
In this paper, density evolution-based construction methods to design good polar codes on impulsive noise channels for single-carrier and multi-carrier systems are proposed and evaluated. For a single-carrier system, the tight bound of the block error probability (BLEP) is derived by applying density evolution and the performance of the proposed construction methods are compared. For the multi-carrier system employing orthogonal frequency-division multiplexing, the accurate BLEP estimation is not feasible so a tight lower bound on the BLEP for polar codes is derived by assuming the noise on each sub-carrier is Gaussian. The results show that the lower bound becomes tighter as the number of carriers increases.
 
\end{abstract}

\begin{keywords}
Impulsive noise, Polar codes, OFDM, density evolution
\end{keywords}

\section{Introduction}
\IEEEPARstart{C}{hannel} polarization, proposed by Arikan, is a method of constructing codes that can achieve the capacity of symmetric binary-input discrete memoryless channels (B-DMCs) \cite{arikan2009channel}. Polar codes, which are constructed using channel polarization, have the advantage of low encoding and decoding complexity. Hence, polar codes are very competitive among capacity-approaching codes such as turbo codes and low-density parity-check (LDPC) codes and very recently, polar codes were adopted in the 5G standard. Arikan proposed a heuristic construction method of polar codes that can achieve the capacity of the binary erasure channel (BEC) only \cite{arikan2008performance}, but alternative construction methods such as density evolution (DE) or Gaussian approximation (GA) based constructions have been proposed to design polar codes specifically for other B-DMCs and additive white Gaussian noise (AWGN) channels \cite{mori2009performance, trifonov2012efficient} with better performance than that achieved by the heuristic method.
\par Most literature on polar codes focuses on the AWGN channel, but there are many applications such as the power-line channel (PLC) and underwater acoustic channels where the received signal is impaired by non-Gaussian impulsive noise. The occurrence of impulsive noise leads to a heavy-tailed probability density function (pdf) for the noise at the receiver, thus the assumption that the noise has a Gaussian distribution is no longer valid. The Bernoulli-Gaussian model, Middleton Class A and B model and symmetric $\alpha$-stable (S$\alpha$S) distributions \cite{ middleton1999non, tsihrintzis1995performance} have been used to approximate the pdf of impulsive noise. In particular, Middleton Class A noise has been investigated by a number of publications, such as \cite{haring2002performance}. It is well-known that orthogonal frequency-division multiplexing (OFDM) is very suitable for the transmission over impulsive noise since the impulses are suppressed after demodulation by spreading them over a large number of carriers. Moreover, with non-linear operations such as clipping or blanking, the effect of impulsive noise can be further reduced \cite{zhidkov2008analysis}.

\par To improve the performance of the system, error correcting codes such as convolutional codes, LDPC codes and turbo codes can be used. And the coded performance on impulsive noise channels has been investigated in many literature \cite{tseng2013robust, mei2017performance}. In particular, LDPC codes have been adopted in the G.hn/G.9960 standard for PLC. Recently, polar codes have been employed to mitigate impulsive noise and shown comparable performance to LDPC codes but with lower complexity \cite{jin2015performance,jin2015performance11,hadi2016polar}. However, the polar codes used in the literature were only constructed using the unoptimized heuristic construction method which leads to a sub-optimal performance. Moreover, after a comprehensive literature review, it appears that the theoretical performance of polar codes has not been studied, which will be addressed in this paper. DE-based construction methods are presented for single-carrier and OFDM systems with impulsive noise. For the single-carrier system, the exact block error probability (BLEP) is derived, but for the OFDM system, only an estimation of the BLEP is given as it is not feasible to derive the exact BLEP.

\section{Background}
\subsection{Channel Model}
We first consider a polar-coded system where the $k$-th received signal $y_{k}$ can be expressed as
\begin{equation}
y_{k}=x_{k}+z_{k},
\end{equation}
where $x_{k}$ is the transmitted signal employing binary phase-shift keying (BPSK) and $x_{k}\in\left\lbrace +1, -1\right\rbrace $, $z_{k}$ is the impulsive noise which is represented by the Middleton Class A noise model. For the real channel, the pdf of $z_{k}$ is given as
\begin{equation} \label{pdf_real}
p_{ZR}(z)=\exp{(-A)}\sum_{m=0}^{\infty}\frac{A^{m}}{m!\sqrt{2\pi\sigma_{m}^{2}}}\exp{\left( -\frac{z^{2}}{2\sigma_{m}^{2}}\right) },
\end{equation}
where $A$ ($A\leq 1$) is the impulsive index which represents the density of impulses in an observation period. The variance $\sigma_{m}^{2}$ is defined as $\sigma_{m}^{2}=\sigma_{g}^{2}\left( \frac{m}{A\Gamma}+1   \right)$,
where $\sigma_{g}^{2}$ is the variance of background Gaussian noise and $\Gamma$ is the background-to-impulsive noise power ratio. For the complex channel, the probability density function (pdf) of $z_{k}$ is defined as
\begin{equation}\label{pdf_com}
p_{Z}(z)=\exp{(-A)}\sum_{m=0}^{\infty}\frac{A^{m}}{m!2\pi\sigma_{m}^{2}}\exp{\left( -\frac{|z|^{2}}{2\sigma_{m}^{2}}\right) }.
\end{equation}

\subsection{Polar Coding}
Channel polarization is a method to construct capacity-approaching codes for symmetric binary-input discrete memoryless channels (B-DMCs). As the length of the codeword $N$ ($N=2^{n}$, $n=1,2,...$) approaches infinity, the polarized channels tend to two extremes: noiseless channels and channels with full noise \cite{arikan2009channel}.
By transmitting information bits on noiseless channels and frozen bits on pure noise channels, polar codes can achieve channel capacity under a successive cancellation (SC) decoder.
\par A binary polar code can be specified by $(N,K,\mathcal{I})$, where $K$ is the message length and $\mathcal{I}\subseteq \left\lbrace 1,2,\cdots, N \right\rbrace $ represents the set of indices of the information bits transmitted. Let $\textbf{u}_{1}^{N}=(u_{1},u_{2},\cdots,u_{N})$ and $\textbf{u}_{i}^{j}=(u_{i},u_{i+1},\cdots,u_{j})$ is a subvector of $\textbf{u}_{1}^{N}$. The codeword $\textbf{c}_{1}^{N}=\textbf{u}_{1}^{N} \textbf{G}_{N}$. We note that the frozen bits are transmitted as zeros and known to both the encoder and the decoder. The generator matrix $\textbf{G}_{N}=\textbf{B}_{N}\textbf{F}_{2}^{\otimes n}$, where $\textbf{B}_{N}$ is the bit-reversal permutation matrix and $\textbf{F}_{2}^{\otimes n}$ is the $n$-th Kronecker power of $\textbf{F}_{2}=
\left[ \begin{matrix}
1 & 0\\ 
1 & 1 \\ 
\end{matrix}\right]. 
$
To decode polar codes, a low complexity SC decoding is used \cite{arikan2009channel}. We denote the $i$-th bit channel as $W_{N}^{(i)}$ with transition probability $p(y_{i}|x_{i})$. In SC decoding, the hard decision of the frozen bits results in a zero. The information bits are decoded sequentially using maximum likelihood (ML) decoding of the bit channel $W_{N}^{i}$. The decoder calculates the log-likelihood ratio (LLR) of the $i$-th bit as
\begin{equation} \label{LLR}
L_{N}^{(i)}(\textbf{y}_{1}^{N},\hat{\textbf{u}}_{1}^{i-1})=\log\frac{p(\textbf{y}_{1}^{N},\hat{\textbf{u}}_{1}^{i-1}|0)}{p(\textbf{y}_{1}^{N},\hat{\textbf{u}}_{1}^{i-1}|1)},
\end{equation}
where $\textbf{y}_{1}^{N}$ is the received sequence with $\textbf{y}_{1}^{N}=(y_{1},y_{2},\cdots,y_{N})$ and $\hat{\textbf{u}}_{1}^{i-1}$ denotes the estimated vector of $\textbf{u}_{1}^{i-1}$. We note that $L_{N}^{(i)}(\textbf{y}_{1}^{N},\hat{\textbf{u}}_{1}^{i-1})$ can be recursively calculated by transforming the update rule in \cite{arikan2009channel} to the log-domain. For information bits, the decision of the $i$-th bit is given as
\begin{equation} \label{decision}
    \hat{u}_{i}=
   \begin{cases}
   0, &\mbox{if $L_{N}^{(i)}(\textbf{y}_{1}^{N},\hat{\textbf{u}}_{1}^{i-1})>0,$}\\
   1, &\mbox{otherwise.}  
   \end{cases}
\end{equation}

\section{Construction and Performance of Polar codes in Single-carrier system with impulsive noise}
In this section, two construction methods for Middleton Class A noise channels will be presented. The first code construction method is based on channel polarization using the Bhattacharyya parameter, $Z(W_{N}^{i})$, for the $i$-th channel. $Z(W_{N}^{i})$ gives an upper bound on the error probability with ML decision and is defined as
\begin{equation}
Z(W_{N}^{(i)})=\sum_{\textbf{y}_{1}^{N},\textbf{u}_{1}^{i-1}}\sqrt{p(\textbf{y}_{1}^{N},\hat{\textbf{u}}_{1}^{i-1}|0)p(\textbf{y}_{1}^{N},\hat{\textbf{u}}_{1}^{i-1}|1)}，
\end{equation}
where $Z(W_{N}^{i})$ for the binary erasure channel (BEC) can be calculated using simple recursion as Eq.4 in \cite{arikan2008performance} and $Z(W_{1}^{(1)})$ is initialised to $1/2$. Then, the bit channels with the $K$ smallest $Z(W_{N}^{(i)})$ are chosen to be the noiseless channels. For Middleton Class A noise channels, $Z(W_{1}^{(1)})$ can be calculated as
\begin{align} \label{recursion}\nonumber
Z(W_{1}^{(1)})&=\int_{-\infty}^{\infty}\sqrt{P(y|x=1)P(y|x=-1)}dy \\ \nonumber
&=\int_{-\infty}^{\infty}\sqrt{\sum_{m=0}^{\infty}\alpha_{m}\exp{\left( -\frac{(y-1)^{2}}{2\sigma_{m}^{2}}\right) }}\times \\
&\qquad \qquad\sqrt{\sum_{m=0}^{\infty}\alpha_{m}\exp{\left( -\frac{(y+1)^{2}}{2\sigma_{m}^{2}}\right) }}dy,
\end{align}
where $\alpha_{m}=\exp(-A)\frac{A^{m}}{m!\sqrt{2\pi\sigma_{m}^{2}}}$. For BEC channel, $Z(W_{1}^{(1)})$ is initialized to $1/2$ which corresponds to 0.05 dB according to \eqref{recursion}, but this may not be the best design-SNR for Middleton Class A noise. Hence the optimized design-SNR should be obtained by simulations. Moreover, $Z(W_{N}^{(i)})$ of Middleton Class A channels is difficult to calculate. To give a tight bound on the BLEP and design better polar codes than those constructed from the heuristic method, a DE-based construction method proposed in \cite{mori2009performance} is used. 
\par To enable DE analysis, first we can prove that the Middleton Class A noise channel is symmetric, as shown below:
\begin{align} \nonumber
P(y|x=1) &= \exp{(-A)}\sum_{m=0}^{\infty}\frac{A^{m}}{m!\sqrt{2\pi\sigma_{m}^{2}}}\exp{\left( -\frac{(y-1)^{2}}{2\sigma_{m}^{2}}\right) } \\ 
&= P(-y|x=-1).
\end{align}

Then, according to \cite{richardson2008modern}, the pdf of the LLR values for any binary memoryless symmetric channel (BMSC) is also symmetric, which enable us to perform DE. Let us denote $a_{N}^{(i)}(y)$ as the pdf of $L_{N}^{(i)}(\textbf{y}_{1}^{N},\textbf{u}_{1}^{i-1})$ when the all-zero codeword is transmitted. The DE of the LLR update in SC decoding is given as
\begin{align} \label{DE} \nonumber
a_{N}^{(2i-1)}(y)&=a_{N/2}^{(i)}(y)\boxtimes a_{N/2}^{(i)}(y), \\
a_{N}^{(2i)}(y)&=a_{N/2}^{(i)}(y)\star a_{N/2}^{(i)}(y),
\end{align}
where $\boxtimes$ and $\star$ denote the convolution in the check and variable node domain, respectively \cite{richardson2008modern}. 
The error probability of the $i$-th bit channel is given as 
\begin{equation}
P_e^{(i)}= \lim_{ \epsilon\rightarrow +0}\left( \int_{-\infty}^{ -\epsilon}d_{N}^{(i)}(x)dx+\frac{1}{2}\int_{- \epsilon}^{+ \epsilon}d_{N}^{(i)}(x)dx \right)  . 
\end{equation}
 $a_{1}^{(1)}(y)$ is the pdf of the initial LLRs $L(y)$ with
\begin{equation} \label{initial}
L(y)=\log\frac{\sum_{m=0}^{\infty}\frac{A^{m}}{m!\sqrt{2\pi\sigma_{m}^{2}}}\exp{\left( -\frac{(y-1)^{2}}{2\sigma_{m}^{2}}\right) }}{\sum_{m=0}^{\infty}\frac{A^{m}}{m!\sqrt{2\pi\sigma_{m}^{2}}}\exp{\left( -\frac{(y+1)^{2}}{2\sigma_{m}^{2}}\right) }}.
\end{equation} 
However, the pdf of \eqref{initial} cannot be evaluated in analytic form and a histogram method is used to find the pdf. 

After DE, the error probability $P_e^{(i)}$ of the decoder for the $i$-th bit channel can be obtained. Then, the DE-based construction method chooses the $K$ channels with smallest $P_e^{(i)}$ to transmit information. With knowledge of $P_e^{(i)}$, the upper bound on BLEP of polar codes is given as \cite{mori2009performance2}
\begin{equation} \label{PB}
P_{B} \approx 1-\prod_{i\in\mathcal{I}}\left( 1-P_e^{(i)} \right)\leq
\sum_{i\in\mathcal{I}}P_e^{(i)}.
 \end{equation}

\section{Asymptotic block-error probability of Polar codes with OFDM on impulsive noise channels}
In this section, we consider the performance of polar codes with OFDM. In a polar-coded OFDM system with $N$ sub-carriers, assuming perfect synchronization and an ideal channel, the received signal $y(k)$ is 
\begin{equation} \label{OFDM_receive}
y_k=\frac{1}{\sqrt{N}}\sum^{N-1}_{n=0}x_{n}\exp\left(j\frac{2\pi nk}{N}\right)+z_k, \quad k=0,1,\cdots,N-1.
\end{equation}  
At the receiver, $x_n$ is recovered by performing an $N$-point discrete Fourier transform (DFT) on the received signal as follows:
\begin{align} \nonumber
Y_{k}&=\frac{1}{\sqrt{N}}\sum^{N-1}_{n=0}y_{n}\exp\left(-j\frac{2\pi nk}{N}\right) \\
&=x_{k}+Z_{k}, \quad k=0,1,\cdots,N-1,
\end{align}
where 
\begin{equation} \label{fre_Z}
Z_{k}=\frac{1}{\sqrt{N}}\sum^{N-1}_{n=0}z_n\exp\left(-j\frac{2\pi nk}{N}\right),\quad k=0,1,\cdots,N-1.
\end{equation}
Unlike the single-carrier system, the DFT spreads the impulsive noise on all sub-carriers in the frequency domain. According to the Central Limit Theorem, $Z_{k}$ can be approximated as a Gaussian random variable if $N$ is large. Hence, the mean and variance of $Z_{k}$ are given as $\mu_{Z}=\mu_{z}=0$ and $ \sigma_{Z}^{2}=N(\sigma_{z}/\sqrt{N})^{2}=\sigma_{z}^{2}$, respectively. The variance $\sigma_{z}^{2}$ of Middleton Class A noise is calculated as
\begin{align} \nonumber
\sigma_{z}^{2}=\int z^{2}p_{Z}(z)dz =\frac{\exp(-A)\sigma_{g}^{2}}{\Gamma}\sum_{m=0}^{\infty}\frac{A^{m}}{m!}\left( \frac{m}{A}+\Gamma\right). 
\end{align}

This is the benefit of multi-carrier modulation in impulsive noise since each sub-channel can be regarded as an AWGN channel. Based on this Gaussian assumption, the pdf of the initial LLR can be given in closed-form as
\begin{equation} \label{ini}
a_{1}^{(1)}(y)=\sqrt{\frac{\sigma_{z}^{2}}{8\pi}}\exp\left(-\frac{\left( y-\frac{2}{\sigma_{z}^2} \right)^{2}\sigma_{z}^{2}  }{8} \right). 
\end{equation}
\par Following the updated equations of DE, the error probability of the decoder can be obtained. Hence the BLEP of polar codes with OFDM can be obtained by using \eqref{PB}. We note that this BLEP is an approximation of the exact BLEP since we assume $Z_{k}$ is Gaussian distributed. The construction of polar codes with OFDM on Middleton Class A noise channels is similar to AWGN channels due to the Gaussian assumption of each sub-band. Unlike the single-carrier system, we cannot obtain the optimal LLRs for decoding when using OFDM and therefore the performance is not optimal. However, to reduce the total energy of impulsive noises and further improve the performance, we consider two different non-linear operations called blanking and clipping that can be applied to the received signal $y_{k}$ in \eqref{OFDM_receive} before the OFDM demodulator to help mitigate the impulsive noise. These non-linear operations are given as \cite{zhidkov2008analysis}
\begin{equation} \label{hp_sf}
    y_{k}^{b}=
   \begin{cases}
   y_k, &\mbox{$|y_k|<T$}\\
   0, &\mbox{$|y_k|\geq T$}  
   \end{cases},
   \qquad 
    y_{k}^{c}=
   \begin{cases}
   y_k, &\mbox{$|y_k|<T$}\\
   Te^{j\arg(y_{k})}, &\mbox{$|y_k|\geq T$} 
   \end{cases} ,  \nonumber
\end{equation}
where $k=0,1,\cdots,N-1$. $y_{k}^{c}$ and $y_{k}^{c}$ are the output of the blanking and clipping non-linearity, respectively and $T$ is the clipping threshold. The performance of polar codes in the single-carrier system and OFDM system will now be compared in the next section.

\section{Results}


\begin{figure}[t] 
\centering
\includegraphics[height=2.3in,width=3.2in]{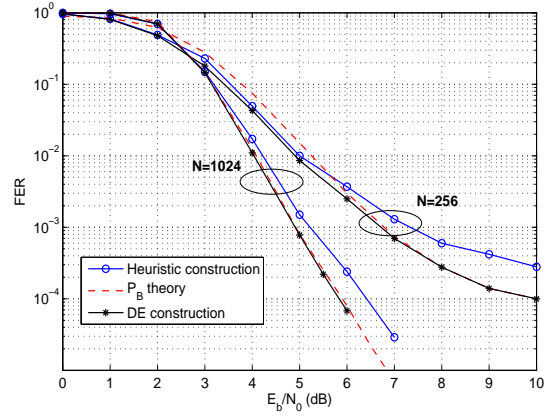}
\caption{Performance of heuristic and DE-based construction methods for the single-carrier system, with $N=1024$ and $\Gamma=0.1, A = 0.1$}
\label{heu_DE}
\end{figure} 

In this section, the theoretical and simulated performance of polar codes for single-carrier and OFDM systems on Middleton Class A noise channels is presented. The parameters of the noise are set to be $\Gamma=0.1$ and $\Gamma=0.3$, $A = 0.1$ which represent strong and moderate impulsive noise. The polar codes we employed are half rate and $N=256, 512, 1024, 4096$. 

\begin{figure}[t] 
\centering
\includegraphics[height=2.3in,width=3.2in]{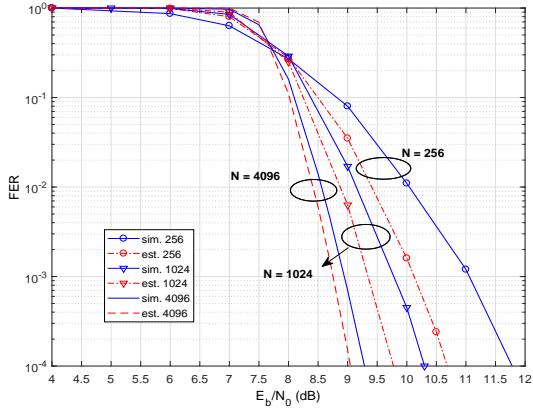}
\caption{Performance of polar codes using OFDM, with block length $N=256,1024,2048$ and $\Gamma=0.3, A = 0.1$}
\label{OFDM}
\end{figure}
\begin{figure}[t] 
\centering
\includegraphics[height=2.2in,width=3.2in]{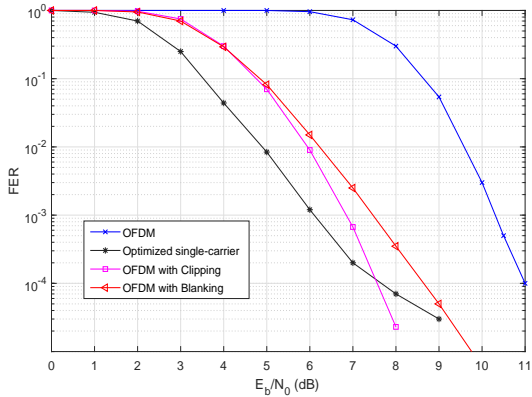}
\caption{Performance comparison of polar codes of single-carrier and OFDM systems, with block length $N=512$ and $\Gamma=0.3, A = 0.1$}
\label{OFDM_nonlinear}
\end{figure}

As shown in Fig. \ref{heu_DE}, the performance of DE construction and heuristic method is compared for different code length in single-carrier system. We note that the design-SNR of heuristic method has already been optimized by experiments. It is shown that the performance of the DE-based construction is better than the heuristic method, which achieves 0.5 dB gain for $N=1024$ at $\text{FER}=10^{-4}$ and greatly lowers the error floor for $N=256$. In addition, the derived theoretical BLEP in \eqref{PB} accurately predicts the simulated performance. Furthermore, we observe that when the block length is short ($N=256$), a severe error floor is present even with optimal LLR values. In Fig. \ref{OFDM}, the performance of polar codes with OFDM are presented. By assuming the received signal after OFDM demodulation is Gaussian, our estimation becomes more accurate as codeword length increases. When $N=4096$, the gap between the estimation and simulated performance is only 0.2 dB at $\text{FER}=10^{-4}$. In Fig. \ref{OFDM_nonlinear}, the performance of the single-carrier and OFDM systems are compared. As shown in Fig. \ref{OFDM_nonlinear}, when the codeword length is not sufficiently long ($N=512$), the polar codes in the single-carrier system exhibit an error floor. However, there is no error floor with OFDM since the impulsive noise is suppressed and spread over all sub-carriers. Interestingly, it is observed that the waterfall performance of OFDM is much worse than single-carrier system since the energy of the impulses is very high. The performance of OFDM can be significantly improved when the clipping or blanking operation is employed and it is shown that the performance is better than the single-carrier system at the low error rate region.


\section{Conclusion}
In this paper, we have investigated two construction methods for polar codes on impulsive noise channels and presented their theoretical performance on these channels. For single carrier modulation, the optimal design-SNR using the heuristic construction method was found for polar codes on Middleton Class A impulsive noise channels and the BLEP performance was evaluated. However, we showed that our DE-based method exhibits a slight performance gain over the heuristic method. With OFDM, an accurate BLEP estimation is not feasible but it was shown that by treating each sub-channel as Gaussian, the performance of polar codes can be well estimated and this becomes more accurate as the block length increases. When the block length is 4096 bits, the gap between estimated and simulated performance is only 0.2 dB. Finally, for the single-carrier system, when the block length is short, a severe error floor is present, but by using OFDM the error floor is eliminated since the impulsive noise is spread over all sub-carriers. Moreover, OFDM with simple non-linear operations can further improve the performance, which demonstrates the advantages of employing OFDM in combination with polar codes to mitigate the effects of impulsive noise.

\small
\bibliographystyle{IEEEtran}
\bibliography{refs}
\end{document}